%% file: apssamp.tex
\documentclass[%
 reprint,
superscriptaddress,
 amsmath,amssymb,
 aps,
pra, longbibliography,
]{revtex4-2}
\usepackage[english]{babel}
\addto\captionsenglish{}
\usepackage{graphicx}
\usepackage{dcolumn}%
\usepackage{bm}%
\usepackage{placeins}

\usepackage[
  acronym,
  toc,
  nonumberlist,
  sort=standard
]{glossaries}
\newglossarystyle{long3col-nogroup-cap}{%
  \setglossarystyle{long3col}%

}

\newacronymstyle{long-short-it}
  {%
    \GlsUseAcrEntryDispStyle{long-short}%
  }{%
    \GlsUseAcrStyleDefs{long-short}%
  }
\setacronymstyle{long-short-it}
\input{acronyms}

\begin{document}

\preprint{APS/123-QED}

\title{Twin-photon generation in a silicon nitride microresonator}

\author{Franz Pacher}
\altaffiliation{These authors contributed equally to this work.}
\affiliation{Max Planck Institute for the Science of Light, Staudtstra{\ss}e 2, 91058 Erlangen, Germany}
\affiliation{Department of Physics, Friedrich-Alexander-Universit\"at Erlangen-N\"urnberg, 91058 Erlangen, Germany}

\author{Haochen Yan}
\altaffiliation{These authors contributed equally to this work.}
\affiliation{Max Planck Institute for the Science of Light, Staudtstra{\ss}e 2, 91058 Erlangen, Germany}
\affiliation{Department of Physics, Friedrich-Alexander-Universit\"at Erlangen-N\"urnberg, 91058 Erlangen, Germany}

\author{Alekhya Ghosh}
\altaffiliation{These authors contributed equally to this work.}
\affiliation{Max Planck Institute for the Science of Light, Staudtstra{\ss}e 2, 91058 Erlangen, Germany}
\affiliation{Department of Physics, Friedrich-Alexander-Universit\"at Erlangen-N\"urnberg, 91058 Erlangen, Germany}

\author{Arghadeep Pal}
\altaffiliation{These authors contributed equally to this work.}
\affiliation{Max Planck Institute for the Science of Light, Staudtstra{\ss}e 2, 91058 Erlangen, Germany}
\affiliation{Department of Physics, Friedrich-Alexander-Universit\"at Erlangen-N\"urnberg, 91058 Erlangen, Germany}

\author{Toby Bi}
\affiliation{Max Planck Institute for the Science of Light, Staudtstra{\ss}e 2, 91058 Erlangen, Germany}
\affiliation{Department of Physics, Friedrich-Alexander-Universit\"at Erlangen-N\"urnberg, 91058 Erlangen, Germany}

\author{Hao Zhang} 
\affiliation{Max Planck Institute for the Science of Light, Staudtstra{\ss}e 2, 91058 Erlangen, Germany}

\author{Lixing You}
\affiliation{Shanghai Key Laboratory of Superconductor Integrated Circuit Technology, Shanghai Institute of Microsystem and Information Technology, Chinese Academy of Sciences, Shanghai 200050, China}

\author{Hao Li}
\affiliation{Shanghai Key Laboratory of Superconductor Integrated Circuit Technology, Shanghai Institute of Microsystem and Information Technology, Chinese Academy of Sciences, Shanghai 200050, China}

\author{Daniela Salvoni}
\affiliation{Photon Technology Italy SRL, Via G. Gigante 174, 80128 Napoli, Italy}

\author{Shuangyou Zhang}
\affiliation{Technical University of Denmark, 2800 Kgs. Lyngby, Denmark} 

\author{Pascal Del'Haye}
\email{pascal.delhaye@mpl.mpg.de}
\affiliation{Max Planck Institute for the Science of Light, Staudtstra{\ss}e 2, 91058 Erlangen, Germany}
\affiliation{Department of Physics, Friedrich-Alexander-Universit\"at Erlangen-N\"urnberg, 91058 Erlangen, Germany}
\date{\today}%

\begin{abstract}
Photonic chips with silicon nitride ($\mathrm{Si_3N_4}$) microring resonators are well established
as heralded single-photon sources, but their operation as
frequency-degenerate twin-photon sources has not previously been
demonstrated. 
Here, we realise a twin-photon source at telecommunication wavelengths in a $\mathrm{Si_3N_4}$
ring microresonator via an inverse \acrfull{fwm} process, 
in which two photons from spectrally distinct pumps are converted into a pair of identical twin photons.
The measurements show a
maximum \acrfull{car} of $5.4\pm0.6$. 
In addition, the microresonator
functions as a heralded
single-photon source through pump-degenerate \acrfull{sfwm}, exhibiting a
spectral purity of $P=0.67\pm0.05$ and a heralded
anti-bunching of $g^{(2)}_h(0)=0.0042\pm0.0015$. 
Together, these results demonstrate both
photon-generation schemes on a single integrated $\mathrm{Si_3N_4}$
platform, highlighting its potential for scalable, tailored quantum
light generation.

\end{abstract}
\maketitle

\section{\label{sec:level1}Introduction}
Non-classical light sources are an essential resource for photonic quantum
technologies, underpinning applications in quantum computing, communication,
metrology and information processing~\cite{obrien_photonic_2009, flamini_photonic_2019}. 
Integrated photonic platforms are particularly attractive in this context, since they
promise the scalability, stability and compact size required to move
quantum optical experiments from bulk-optics laboratory setups towards
deployable devices~\cite{wang2020integrated, pelucchi2022potential,
caspani_integrated_2017, signorini_-chip_2020}. Among these platforms,
silicon nitride ($\mathrm{Si_3N_4}$) has emerged as a leading candidate for
integrated nonlinear and quantum photonics, combining ultralow propagation
loss, a broad transparency window, an absence of two-photon absorption at
telecommunication wavelengths, and full compatibility with \acrshort{cmos}
fabrication~\cite{buzaverov_silicon_2024, zhang_lowtemperature_2024,
xiang2022silicon, blumenthal2018silicon, ghosh2026fourth,
stern_battery-operated_2018}. 
Owing to the relatively high $\chi^{(3)}$
susceptibility of $\mathrm{Si_3N_4}$, high-$Q$ ring microresonators support
efficient nonlinear interactions ranging from broadband frequency comb and
soliton formation~\cite{kippenberg_dissipative_2018,
gaeta_photonic-chip-based_2019, pal2026hybrid} to symmetry breaking~\cite{white2023integrated, zhang2025integrated, trinchao2026color}, as well as correlated photon-pair
generation via spontaneous four-wave mixing (\acrshort{sfwm})~\cite{wang_progress_2024,reimer_integrated_2014, camacho_entangled_2012}.
 
\acrshort{sfwm} in a ring microresonator can be operated in two
complementary configurations. 
Pump-degenerate \acrshort{sfwm},
in which a single cavity resonance is pumped, generates
frequency-distinct signal and idler photons and forms the basis of
well-established heralded single-photon sources~\cite{fortsch_versatile_2013,
reimer_integrated_2014}. 

\begin{figure}[!htb]
    \centering
    \includegraphics{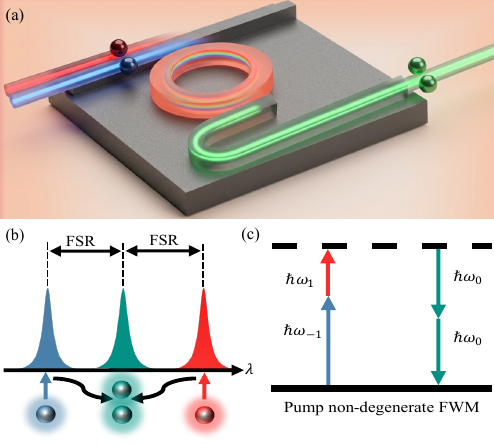} 
    \caption{Twin-photon generation in a ring microresonator.
    (a) Schematic illustration of twin-photon generation through dual-pumped ``inverse" \acrfull{fwm} in a ring microresonator. 
    (b) Cavity resonance spectrum showing the two pump modes (red and blue) positioned symmetrically about the central signal/idler mode (green), with a separation of one free spectral range (FSR) between each pump and the central mode.
    (c) Energy-level diagram: absorption of two photons with energies $\hbar\omega_{-1}$ and $\hbar\omega_{1}$ generates two degenerate photons with energies $\hbar\omega_0$, satisfying $\omega_{-1}+\omega_{1}=2\omega_0$.
    }
    \label{fig:concept_figure}
\end{figure}
Alternatively, pump non-degenerate \acrshort{sfwm}, in which two cavity resonances are pumped simultaneously, generates frequency-\emph{degenerate} photon pairs: 
the resulting twin photons are
indistinguishable in every degree of freedom~\cite{rogers_twin_2015} and,
at low pump powers, 
the leading non-vacuum contribution is the two-photon Fock
component of a single-mode squeezed vacuum. 
Figure~\ref{fig:concept_figure} depicts the pump non-degenerate
\acrshort{sfwm} process schematically.
This dual-pumped configuration realises the photon-counting regime of an integrated single-mode squeezer, which is an
elementary resource for a broad class of continuous-variable and multi-photon quantum protocols~\cite{larsen_integrated_2025}.
Despite prior demonstrations on other platforms~\cite{rogers_twin_2015, azzini_ultra-low_2012, guo_telecom-band_2014, chen_fiber-based_2006, fan_generation_2005, engin_photon_2013, he_degenerate_2014},
a frequency-degenerate twin-photon source remains unexplored in an integrated $\mathrm{Si_3N_4}$ device.
Here we demonstrate that a single $\mathrm{Si_3N_4}$ ring microresonator simultaneously supports two quantum light source functionalities: a frequency-degenerate twin-photon source at telecommunication wavelengths through pump non-degenerate SFWM, and a high-purity heralded single-photon source through pump-degenerate \acrshort{sfwm}. 
These two functionalities can be accessed simply by changing the pumping scheme. 
The generated photons are filtered and then detected using highly efficient superconducting nanowire single-photon detectors
(\acrshortpl{snspd})~\cite{you_superconducting_2020}. 
To our knowledge, this establishes the first frequency-degenerate
twin-photon source in $\mathrm{Si_3N_4}$, together with the first
demonstration of both photon-generation schemes in the same device.

\section{Results}
\subsection{Frequency-degenerate twin-photon source} 

\begin{figure*}[t!]
    \centering
    \includegraphics{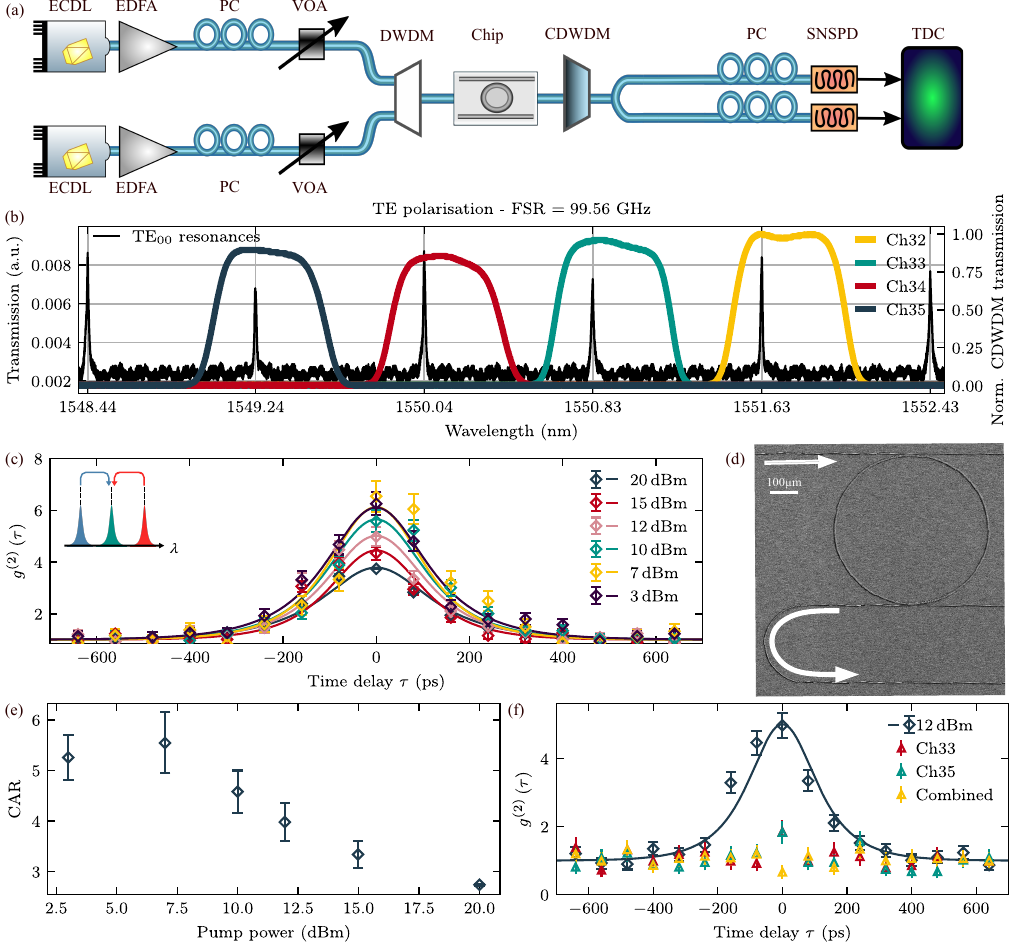}%
    \caption{Pump non-degenerate \acrshort{sfwm} in a ring microresonator. 
    (a) Experimental setup for twin-photon generation using pump non-degenerate \acrshort{sfwm}.
    \acrshort{ecdl}: \acrlong{ecdl}; \acrshort{edfa}: \acrlong{edfa}; \acrshort{pc}: \acrlong{pc};  \acrshort{voa}: \acrlong{voa}; \acrshort{dwdm}: \acrlong{dwdm}; \acrshort{cdwdm}: \acrlong{cdwdm}; \acrshort{snspd}: \acrlong{snspd}; \acrshort{tdc}: \acrlong{tdc}.
    (b) Alignment of the ring microresonator spectrum with the ITU-T 
    $100\,\mathrm{GHz}$ \acrshort{cdwdm} channel grid. The measured cavity transmission
    in \acrshort{te} polarisation (dark trace) exhibits a resonance
    approximately every one \acrshort{fsr} ($99.56\,\mathrm{GHz}$), each of
    which falls within a distinct \acrshort{cdwdm} channel passband (coloured
    bands, labelled by ITU channel number).
    (c) $g^{(2)}$-correlation measurements of the twin-photon source for multiple input powers, together with fits to the data (solid lines). For more information on the fits, see Appendix~\ref{app:g2_fits}.
    Inset: Schematic of pump non-degenerate \acrshort{sfwm} at resonance wavelengths of the device.
    (d) \Acrfull{sem} picture of the resonator and waveguides. The white arrows indicate the direction of the guided modes to and from the resonator.
    (e) Extracted CAR value of all the $g^{(2)}$-data recorded [shown in (c)] for different input powers.
    (f) $g^{(2)}$-correlation measurement at $12\,\mathrm{dBm}$, with a fit to the data (solid line), together with background measurements.
    The blue data points stem from pump non-degenerate \acrshort{sfwm}, while the red and green data points are the result of pump-degenerate \acrshort{sfwm}. 
    In the latter case only one of the two pump resonances was driven, with the other pump switched off. 
    The yellow data set combines the results of the red and the green data sets. 
    Error bars in $g^{(2)}$-histograms are calculated assuming Poissonian statistics of the recorded 
    data points.
    }
    \label{fig:Dual_pumping}
\end{figure*}

The frequency-degenerate twin-photon source is implemented by pumping the
ring resonator bichromatically, with two lasers tuned to the cavity
resonances aligned with Ch~33 and~35 of a commercial $100\,\mathrm{GHz}$ \acrfull{cdwdm}, acting as a filter, 
separated by two \acrshortpl{fsr}. 
This dual-pumping scheme drives a pump non-degenerate \acrshort{sfwm} process
that generates frequency-degenerate signal and idler photons in the
central resonance (Ch~34). 
The combined pump power is kept
sufficiently low, such that the leading non-vacuum contribution is the two-photon Fock component of the corresponding single-mode squeezed vacuum. 
To characterise the twin-photon generation process, $g^{(2)}$-correlation measurements are performed on the central resonance channel.
The full optical setup is shown in Fig.~\ref{fig:Dual_pumping}(a).
The \acrshort{cdwdm} transmission windows together with the cavity resonances are shown in Fig.~\ref{fig:Dual_pumping}(b), highlighting the filtering between different frequency modes.
Repeating the $g^{(2)}$-correlation measurements while varying the total pump power (with equal power in the two pumps) reveals a clear power dependence of $g^{(2)}(0)$, which increases with decreasing pump power [Fig.~\ref{fig:Dual_pumping}(c)]. 
Extracting
the \acrfull{car} from each $g^{(2)}(0)$ value
as $\mathrm{CAR}=g^{(2)}(0)-1$
[Fig.~\ref{fig:Dual_pumping}(e)] reproduces the expected inverse
dependence of the \acrshort{car} on pump power, reaching a maximum of
$\mathrm{CAR}=5.4\pm0.6$ at a combined pump power of $7\,\mathrm{dBm}$;
below this power, no statistically significant further increase is
observed~\footnote{
Note that the 50/50 splitting of degenerate twin photons reduces the measured \acrshort{car} by a factor of two relative to a non-degenerate cross-correlation measurement of the same generation rate, so this figure is not directly comparable in magnitude to the heralded-source \acrshort{car} reported in Sec.~\ref{sc:heralded_source}.}.
Figure~\ref{fig:Dual_pumping}(d) shows a \acrfull{sem} image of the device, with arrows indicating the direction of travel for the guided modes in the waveguides to and from the resonator.
To confirm that pump non-degenerate \acrshort{sfwm} is the dominant
contribution, we compare the $g^{(2)}$-correlation recorded under
dual-pumping against a background measurement in which each pump operates
individually, in analogy to the work of Rogers
\textit{et~al.}~\cite{rogers_twin_2015}. 
As shown [Fig.~\ref{fig:Dual_pumping}(f)], the dual-pumped configuration produces a
strong bunching peak with $g^{(2)}(0)>2$
at zero delay, whereas the single-pump measurements
yield
only the self-correlation results for signal or idler modes.
The significantly higher values of $g^{(2)}(0)$ in the pump non-degenerate \acrshort{sfwm} case confirm the generation
of frequency-degenerate photon pairs against the pump-degenerate \acrshort{sfwm}.

\begin{figure*}[!ht]
\includegraphics{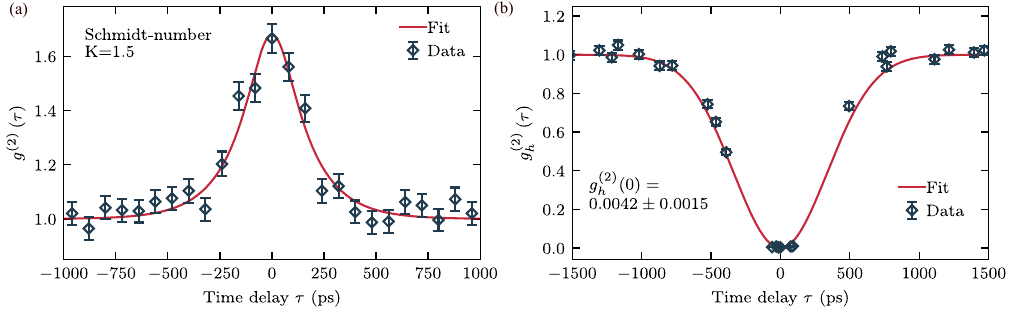}
\caption{Characterisation of the pump-degenerate \acrshort{sfwm}-based heralded single-photon source. 
(a) $g^{(2)}$-self-correlation measurement of the signal mode (Ch~33), with a fit to the data. The self-correlation exhibits a bunching peak with $g^{(2)}(0)=1.67\pm0.05$, corresponding to a Schmidt number of $K=1.49\pm0.12$ ($\approx1.5$ effective modes). 
(b) Heralded $g^{(2)}_h(\tau)$ autocorrelation measurement, with a fit to the data, showing a pronounced anti-bunching dip with $g^{(2)}_h(0)=0.0042\pm0.0015$, well below the single-photon threshold. 
Error bars are calculated assuming Poissonian statistics of the recorded coincidence counts.}
\label{fig:single_photon_purity_heraldedg2}
\end{figure*}


\subsection{Heralded single-photon source}\label{sc:heralded_source}

The same ring resonator, pumped at the resonance aligned with Ch~34 of the \acrshort{cdwdm}, 
additionally functions as a heralded
single-photon source through pump-degenerate \acrshort{sfwm}, generating
frequency-distinct signal and idler photons in the resonances aligned with Ch~33 and
Ch~35, respectively. 
The source is characterised with
respect to its spectral purity, 
the strength of its photon-pair correlations via CAR, 
and the single-photon character of its heralded output. 
All three quantities
are extracted from $g^{(2)}$-correlation measurements recorded with the
\acrshortpl{snspd} and the time-tagging electronics, using the setup
shown in Fig.~\ref{fig:setup_heralding} of the Appendix.

Figure~\ref{fig:single_photon_purity_heraldedg2}(a) shows the measured
spectral purity, assessed through a $g^{(2)}$-self-correlation measurement
of the signal mode (Ch~33), using the setup shown in Fig.~\ref{fig:setup_heralding}(b). 
The signal mode is split by a $3$-dB
fibre coupler, and the coincidences between the two arms are recorded.
The histogram exhibits a bunching peak at zero relative delay with
$g^{(2)}(0)=1.67\pm0.05$, consistent with 
the thermal statistics of the multimode signal field
of the biphoton state.

The value of $g^{(2)}(0)$ is
related to the Schmidt number $K$ by
\begin{equation}\label{eq:Schmidt_number}
    K=\frac{1}{g^{(2)}(0)-1}\,,
\end{equation}
which quantifies the effective number of frequency-mode pairs contributing
to the joint signal--idler state. From the measured $g^{(2)}(0)$ we
obtain a Schmidt number of $K=1.49\pm0.12$ and a purity of $P=1/K=0.67\pm0.05$, indicating approximately $1.5$ effective modes in the
Schmidt decomposition. 
The residual deviation from $K=1$ is attributed to
incomplete spectral isolation of a single signal--idler mode pair after filtering.

The strength of the photon-pair correlations is assessed through a
$g^{(2)}$-cross-correlation measurement, using the setup shown in Fig.~\ref{fig:setup_heralding}(d), between the signal (Ch~33)
and idler (Ch~35) modes while pumping Ch~34. The
cross-correlation exhibits a clear bunching peak at zero delay, from
which the \acrshort{car} is obtained as before. 
Repeating the measurement across a range of
pump powers reproduces the expected inverse dependence of the
\acrshort{car} on pump power, reaching a maximum of $\mathrm{CAR}=9.1$ at
the lowest power measured. An example of a cross-correlation histogram [Fig.~\ref{fig:app_heralding_Car}(a)]
and the extracted \acrshort{car}-versus-power data [Fig.~\ref{fig:app_heralding_Car}(b)] are provided in 
Appendix~\ref{app:supporting_heralding}.

Finally, the source is operated as a heralded single-photon source using
the scheme shown in Fig.~\ref{fig:setup_heralding}(c). The idler photon
(Ch~35) serves as the herald, while the signal mode (Ch~33) is
split by a $3$-dB coupler and sent to two detectors. 
Conditioning the detection of a signal photon in either splitter arm on the detection of
the idler herald exploits the photon-number correlations between the two
modes: heralding significantly reduces the probability of detecting zero photons and 
recovers the sub-Poissonian statistics of the heralded mode, providing
probabilistic single photons. 
The resulting heralded autocorrelation $g^{(2)}_h(\tau)$ shows a
strong anti-bunching dip at zero relative delay
[Fig.~\ref{fig:single_photon_purity_heraldedg2}(b)]. 
The measured value $g^{(2)}_h(0)=0.0042 \pm 0.0015$ lies far below the $g^{(2)}_h(0)<0.5$
threshold for single-photon behaviour, and is among the lowest
reported for integrated microresonator sources~\cite{ma_silicon_2017,
reimer_integrated_2014, reimer_cross-polarized_2015, guo_parametric_2016,
fan_multiwavelength_2023, wu_integrated_2021}, demonstrating high-fidelity
single-photon emission.

\section{Conclusion}

We have demonstrated a frequency-degenerate twin-photon source in an
integrated $\mathrm{Si_3N_4}$ ring microresonator, based on a pump non-degenerate \acrshort{sfwm} process reaching a maximum coincidence-to-accidental ratio of $\mathrm{CAR}=5.4\pm0.6$ at telecommunication wavelengths. 
As far as we are aware, this constitutes the first
frequency-degenerate twin-photon source implemented on this platform.
Via pump-degenerate \acrshort{sfwm}, the same resonator functions as
a high-purity heralded single-photon
source with a spectral purity of $P=0.67\pm0.05$ and a
heralded antibunching value of $g^{(2)}_h(0)=0.0042\pm0.0015$, 
among the lowest reported for integrated microresonator sources.
Together with the twin-photon operation, this provides the first demonstration of both
photon-generation schemes within a single integrated architecture on this
platform.

Higher-quality-factor resonators and improved
long-term pump-laser stability would further increase the \acrshort{car} by enhancing the pair generation rate compared to the
pump-power-dependent background contributions.

Additionally, the dual-pumped configuration
realises the low-gain limit of an integrated single-mode squeezer, which is a critical component for Gaussian boson
sampling~\cite{hamilton_gaussian_2017, madsen_quantum_2022} and continuous-variable quantum
computation~\cite{larsen_integrated_2025}, for which single-mode squeezing~\cite{ulanov_quadrature_2025} 
is required for compatibility with photon-number-resolving readout~\cite{vaidya_broadband_2020}. 
In the discrete-variable regime, the frequency-degenerate twin-photon generation demonstrated here
provides a building block for path-entangled NOON-state generation~\cite{silverstone_-chip_2014}.
The demonstration of both frequency-degenerate twin-photon and heralded single-photon operation
within a single $\mathrm{Si_3N_4}$ microresonator therefore positions this
platform as a bridge between discrete- and continuous-variable integrated
quantum photonics.


\section*{Author Contributions}
FP, HY, AG and AP contributed equally to this work. 
FP and HY conducted the experiments with support from AP and AG. FP analysed the data. FP and AP designed the chip. AG and HY fabricated the sample with support from TB. HY, AP, AG, HZ, and SZ contributed to early-stage scientific discussions and experiments.
All authors discussed the results. FP wrote the manuscript with help from HY, AG, AP and PD. 
LY and HL were primarily responsible for SNSPD fabrication.
DS provided technical services related to SNSPD operation and usage.
PD supervised the project.

\begin{acknowledgments}
The authors would like to thank Irina Harder, Katrin Ludwig, Florentina Gannott, Alexander Gumann, Eduard Butzen, and Heike Schröter-Hohmann from the Technology Development and Service Group for Nanofabrication (MPL, Erlangen). The authors also thank Maria Chekhova (MPL, Erlangen), Raktim Haldar (IIT Bhubaneswar, Bhubaneswar), and Nicolas Joly (MPL, Erlangen) for valuable discussions.
\end{acknowledgments}

\section*{Funding}
This work was supported by 
MQV Project TeQSiC, the German Federal Ministry of Research, Technology and Space, Quantum Systems, 13N17314, 13N17342, the Max Planck Society, and the Max Planck School of Photonics. SZ acknowledges support from Deutsche Forschungsgemeinschaft project 541267874.

\appendix
\section{Methods}\label{app:methods}
\subsection{Experimental setups}\label{app:setup}
\begin{figure*}[!htb]
    \centering
    \includegraphics{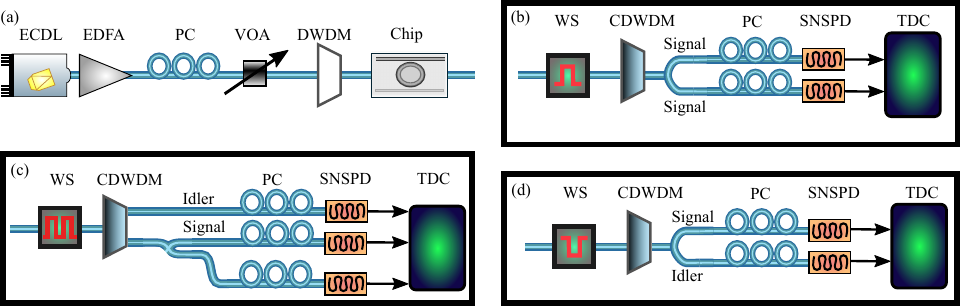} 
    \caption{ Experimental setups used for the pump-degenerate \acrshort{sfwm} heralded photon-source measurements.
    (a) Setup for the generation of a biphoton state through \acrshort{sfwm} in the ring microresonator.
    (b) Setup for measuring the self-correlation between the split signal mode.
    (c) Setup for measuring the heralded autocorrelation between a signal and idler pair.
    (d) Setup for measuring the cross-correlation between a signal and idler mode pair.
    \acrshort{ecdl}: external cavity diode laser; \acrshort{edfa}: erbium-doped fibre amplifier; \acrshort{pc}: polarisation controller; \acrshort{voa}: \acrlong{voa}; \acrshort{dwdm}: \acrlong{dwdm}; \acrshort{ws}: programmable \acrlong{ws}; \acrshort{cdwdm}:  \acrlong{cdwdm}; \acrshort{snspd}: \acrlong{snspd}; \acrshort{tdc}: \acrlong{tdc}.}
    \label{fig:setup_heralding}
\end{figure*}

The optical setups used for the three measurement schemes of the heralded single-photon source are shown in Fig.~\ref{fig:setup_heralding}; the dual-pumped correlation measurement for the twin-photon source is shown in Fig.~\ref{fig:Dual_pumping}(a). 
In all cases, one or two \acrlongpl{ecdl} are amplified with \acrlongpl{edfa} and tuned onto the relevant cavity resonance(s). 
The \acrfullpl{dwdm} were used to reject \acrlong{ase} noise at the chip input,
suppress residual pump light at the chip output, and isolate the individual signal, idler, and pump channels from one another before detection.
In the heralded single-photon source measurements,
\acrshort{dwdm} and \acrshort{cdwdm} filtering stages were supplemented by a programmable \acrlong{ws} for fine spectral selection. 
Due to their shared $100\,\mathrm{GHz}$ ITU-T channel grid, the \acrshort{cdwdm} and \acrshort{dwdm} can be cascaded for filtering.
Light is coupled in and out of the chip with the help of lensed fibres.
A polarisation controller and \acrlong{voa} are used to optimise coupling into the \acrshort{te} mode of the chip and to control the on-chip pump power, respectively. Polarisation controllers were also used to optimise the polarisation of the outcoupled light for detection with the \acrshortpl{snspd}.

The filtered signal and idler photons are detected using \acrshortpl{snspd} from Photon Technology 
with a manufacturer-specified timing jitter of $\approx 80\,\mathrm{ps}$ (\acrshort{fwhm}). Detection events were time-tagged using a 
time-to-digital converter (\acrshort{tdc}) with picosecond timing resolution, from which all $g^{(2)}$-correlation histograms presented in the main text were constructed via arrival-time histograms between the relevant detector channels.

\subsection{Device characterisation}\label{app:device}

\begin{figure*}
    \centering
    \includegraphics{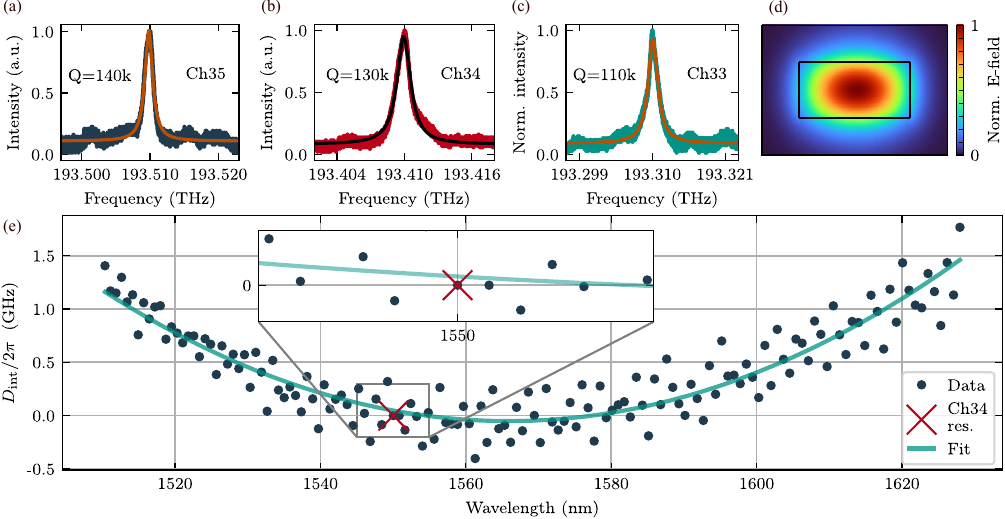}
    \caption{ Characterisation of the ring microresonator device.
    Cavity resonances aligned with Ch~35~(a), Ch~34~(b), and Ch~33~(c) with loaded quality factors of $Q=140\,\mathrm{k}$, $Q=130\,\mathrm{k}$, and $Q=110\,\mathrm{k}$, respectively.
(d) Simulated normalised electric field distribution of the resonator $\mathrm{TE}_{00}$-mode at $1550\,\mathrm{nm}$.
(e) Integrated dispersion of the resonator $\mathrm{TE}_{00}$ around $1550\,\mathrm{nm}$, with quadratic polynomial fit to the data. The dispersion is found to be anomalous, with $D_2/2\pi=264\,\mathrm{kHz}$. The location of the Ch~34 resonance, at $1550\,\mathrm{nm}$, is indicated in the zoomed inset.
}
    \label{fig:dwdm_alignment}
\end{figure*}

The photon sources were implemented on an integrated $\mathrm{Si_3N_4}$ add-drop ring microresonator [Fig.~\ref{fig:Dual_pumping}(d)], chosen for its strong pump rejection at the drop port and low intracavity loss. 
The ring radius of $229.6\,\mathrm{\mu m}$ was chosen to target a \acrshort{fsr} of $100\,\mathrm{GHz}$, 
matching the $100\,\mathrm{GHz}$ channel spacing of the \acrshort{dwdm} filters used in the experiment; the measured average \acrshort{fsr} of \acrshort{te}-polarised modes of
the fabricated device is $99.56\,\mathrm{GHz}$ [Fig.~\ref{fig:Dual_pumping}(b)]. 
This deliberate matching places each cavity resonance within a distinct \acrshort{dwdm} channel,
so that the pump, signal, and idler photons emerge in separate
spectral channels with strong suppression. 
Throughout this work, cavity
resonances are labelled by the ITU channel with which they are aligned:
Ch~34 is aligned with the pump resonance ($1550.04\,\mathrm{nm}$) for
the heralded single-photon source and the degenerate signal/idler
resonance for the twin-photon source, while the two adjacent
resonances, aligned with Ch~35 ($1549.24\,\mathrm{nm}$) and Ch~33 ($1550.83\,\mathrm{nm}$), serve as signal/idler for the heralded source
and as pump resonances for the twin-photon source. 
The alignment of the cavity spectrum with the \acrshort{cdwdm} channel grid, shown in Fig.~\ref{fig:Dual_pumping}(b), 
allows most (for twin-photon generation all) pump-rejection and signal/idler separation to be performed
with off-the-shelf telecom \acrshort{dwdm} components.
The resonances are shown in Fig.~\ref{fig:dwdm_alignment}(a--c) and exhibit loaded quality factors of $Q\approx10^{5}$.
The waveguide core has a cross-section of $1.8\,\mu\mathrm{m}\times700\,\mathrm{nm}$, surrounded by $6\,\mu\mathrm{m}$ of $\mathrm{SiO_2}$, supporting a well-confined fundamental \acrshort{te}-mode [Fig.~\ref{fig:dwdm_alignment}(d)] with anomalous dispersion near the pump wavelengths [Fig.~\ref{fig:dwdm_alignment}(e)].

\subsection{Analytical expressions for $g^{(2)}$-correlation measurements}\label{app:g2_fits}

To account for the influence of finite timing jitter of the \acrshort{snspd} detectors in the $g^{(2)}$-correlation timing histograms, we convolve the functional shape underlying a standard $g^{(2)}$-correlation measurement,
\begin{equation}\label{app_eq:g2_convo_pure_shape}
    f(\tau;\tau_c) = e^{-|\tau|/\tau_c}\,,
\end{equation}
where $\tau$ is the relative delay between coincidences and
$\tau_c$ is the coherence time of the \acrshort{sfwm} process, dictated by the cavity linewidth, with a Gaussian function,
following the treatment by Guo et al.~\cite{guo_parametric_2016}. 
The measured $g^{(2)}(\tau)$ is related to the convolved shape $\mathcal{I}(\tau)$ [Eq.~\eqref{app_eq:convo_integral}] via 
an offset and prefactor ($A$), $g^{(2)}(\tau) = g^{(2)}(\tau=\infty) \pm A\,\mathcal{I}(\tau)/\mathcal{I}(0)$, 
which are  
unaffected by the convolution. 
The prefactor $A$ depends on the generation rate and coherence time of the process, and the type of $g^{(2)}$-measurement.
The Gaussian is described by a second-order moment $\tau_w^2$ given by the quadrature-added jitter times $\tau_{j_n}$ of the $n$ detectors used in a given measurement,
and the bin width $\tau_b$ of the constructed histogram, following
\begin{equation}\label{app_eq:Jitter_width_Gaussian_n_detectors}
    \tau_w^2 = \left( \frac{\tau_b}{2}\right)^2 + \sum_n \tau_{j_n}^2 \;.
\end{equation}
Resulting in a Gaussian given by
\begin{equation}\label{app_eq:Gaussian_Kernel}
    G(\tau,\tau_w)=\frac{1}{\sqrt{2\pi}\,\tau_w} \exp{\left( -\frac{\tau^2}{2\tau_w^2} \right)}\;.
\end{equation}
The convolution of Eq.~\eqref{app_eq:g2_convo_pure_shape} and Eq.~\eqref{app_eq:Gaussian_Kernel}, as a function of the relative delay $\tau$, is defined by
\begin{widetext}
\begin{equation}\label{app_eq:convolution_definition}
\begin{aligned}
     f(\tau;\tau_c) \ast G(\tau,\tau_w)
     \equiv (f\ast G)(\tau) = \int_{-\infty}^{\infty} f(t;\tau_c)\, G(\tau-t,\tau_w) \;dt \,,
\end{aligned}
\end{equation}
which can be evaluated as
\begin{equation}\label{app_eq:convo_integral}
\begin{aligned}
    \mathcal{I}(\tau) \equiv (f\ast G)(\tau)
    =\int_{-\infty}^{\infty}  e^{-|t|/\tau_c} \, \frac{1}{\sqrt{2\pi}\,\tau_w}\, e^{-(\tau-t)^2/2\tau_w^2} \;dt = J_-(\tau) + J_+(\tau) \;,
\end{aligned}
\end{equation}
with
\begin{equation}\label{app_eq:split_integral}
\begin{aligned}
    J_+(\tau)= \int_0^{\infty} e^{-t/\tau_c}  \frac{1}{\sqrt{2\pi}\,\tau_w} e^{-(\tau-t)^2/2\tau_w^2} \;dt\,, \quad
    J_-(\tau)= \int_{-\infty}^0 e^{+t/\tau_c}  \frac{1}{\sqrt{2\pi}\,\tau_w} e^{-(\tau-t)^2/2\tau_w^2} \;dt\;.
\end{aligned}
\end{equation}
Performing the integrals yields
\begin{equation}\label{app_eq:I_integral_solution}
    J_-(\tau) + J_+ (\tau) = \frac{1}{2}\,e^{\frac{\tau_w^2}{2\tau_c^2}}
    \Biggl\{
    \left[ 1 -  \mathrm{erf}\!\left( \frac{\tau + \frac{\tau_w^2}{\tau_c}}{\sqrt{2}\,\tau_w} \right)  \right] e^{+\tau/\tau_c}
    +
    \left[1+ \mathrm{erf}\!\left( \frac{\tau - \frac{\tau_w^2}{\tau_c}}{\sqrt{2}\,\tau_w} \right)  \right] e^{-\tau/\tau_c}
    \Biggr\}\;,
\end{equation}
\end{widetext}
which after normalisation by $\mathcal{I}(\tau=0)= e^{\frac{\tau_w^2}{2\tau_c^2}}\left[1-\mathrm{erf}\!\left(  \frac{\tau_w}{\sqrt{2}\,\tau_c}\right) \right]$ allows fits to the $g^{(2)}$-data that account for detector jitter times comparable to the coherence time $\tau_c$.
For the heralded autocorrelation $g^{(2)}_h(\tau)$ measurement, $\tau_w$ was left as a free parameter, and the excess broadening is attributed to the finite heralding window and the four-fold coincidence construction.
These fits to the $g^{(2)}$-data serve to confirm visually that the measured histograms are consistent with the expected functional form; all quantitative results reported in the main text are obtained directly from the coincidence counts at the relevant time delays, independent of this fit.

\section{Supporting Data: Heralded Single-Photon Source}\label{app:supporting_heralding}

\begin{figure*}[!htb]
    \centering
    \includegraphics{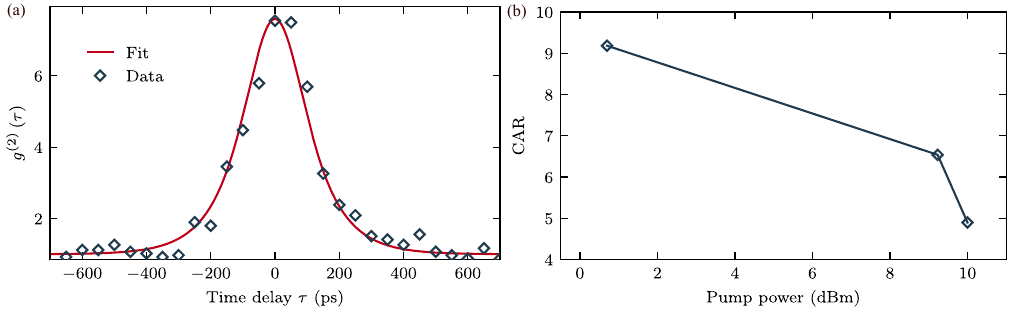} 
    \caption{
    (a) $g^{(2)}$-cross-correlation measurement between the signal mode (Ch~33) and idler mode (Ch~35)  at $9.23\,\mathrm{dBm}$ pump power, reaching $g^{(2)}(0)=7.5$.
    (b) Coincidence-to-accidental ratio (CAR) of the heralded single-photon source, extracted from $g^{(2)}$-cross-correlation measurements, plotted against the pump power onto the chip. The CAR increases with decreasing pump power, reaching a maximum of $9.1$.
    Statistical uncertainties are not shown for these measurements.
    }
    \label{fig:app_heralding_Car}
\end{figure*}

An example of a $g^{(2)}$-cross-correlation measurement of the heralded single-photon source, recorded at a pump power of $9.23\,\mathrm{dBm}$, is shown in Fig.~\ref{fig:app_heralding_Car}(a), together with a fit to the data (red). 
The corresponding \acrshort{car}-versus-power relation, obtained from a set of such cross-correlation measurements, is shown in Fig.~\ref{fig:app_heralding_Car}(b) and is consistent with the expected increase in \acrshort{car} with decreasing pump power, reaching the maximum value reported in the main text at the lowest power measured.

\bibliographystyle{apsrev4-2}
\bibliography{apssamp}%

\end{document}

%% file: acronyms.tex
\newacronym{fwm}{FWM}{four-wave mixing}
\newacronym{sfwm}{SFWM}{spontaneous four-wave mixing}
\newacronym{fsr}{FSR}{free spectral range}
\newacronym{cmos}{CMOS}{complementary metal-oxide semiconductor}
\newacronym{car}{CAR}{coincidence-to-accidental ratio}
\newacronym{tdc}{TDC}{time-to-digital converter}
\newacronym{fwhm}{FWHM}{full width at half maximum}
\newacronym{edfa}{EDFA}{erbium-doped fibre amplifier}
\newacronym{snspd}{SNSPD}{superconducting nanowire single-photon detector}
\newacronym{te}{TE}{transverse electric}
\newacronym{dwdm}{DWDM}{dense wavelength division multiplexer}
\newacronym{cdwdm}{CDWDM}{cascaded dense wavelength division multiplexer} 
\newacronym{ase}{ASE}{amplified spontaneous emission}
\newacronym{ecdl}{ECDL}{external cavity diode laser}
\newacronym{voa}{VOA}{variable optical attenuator}
\newacronym{pc}{PC}{polarisation controller}
\newacronym{ws}{WS}{waveshaper}
\newacronym{sem}{SEM}{scanning electron microscope}